\documentclass[10pt,twocolumn,letterpaper]{article}

\usepackage[final]{iccv}      
\usepackage{algorithm}
\usepackage{algpseudocode}
\usepackage{multicol, multirow}
\usepackage{pifont}
\usepackage{makecell} 
\usepackage{colortbl}
\newcommand{\figref}[1]{Fig.~\ref{#1}}
\newcommand{\tabref}[1]{Tab.~\ref{#1}}
\newcommand{\secref}[1]{Sec.~\ref{#1}}

\definecolor{iccvblue}{rgb}{0.21,0.49,0.74}
\usepackage[pagebackref,breaklinks,colorlinks,allcolors=iccvblue]{hyperref}

\def\paperID{5020} 
\def\confName{ICCV}
\def\confYear{2025}

\title{Degradation Alchemy: Self-Supervised Unknown-to-Known Transformation for Blind Hyperspectral Image Fusion}

\author{He Huang$^1$ \quad Yong Chen$^2$ \quad Yujun Guo$^1$ \quad Wei He$^1$\thanks{Corresponding author.} \\ \\
$^1$ LIESMARS, Wuhan University, Wuhan, China\\
$^2$ School of Computer and Information Engineering, Jiangxi Normal University, Nanchang, China \\
{\tt\small huang\_he@whu.edu.cn} \quad {\tt\small chenyong1872008@163.com}
\quad {\tt\small yujunguo@whu.edu.cn} \quad {\tt\small weihe1990@whu.edu.cn}
}
\begin{document}
\maketitle
\begin{abstract}
      Hyperspectral image (HSI) fusion is an efficient technique that combines
      low-resolution HSI (LR-HSI) and high-resolution multispectral images (HR-MSI)
      to generate high-resolution HSI (HR-HSI). 
      Existing supervised learning methods (SLMs) can yield promising results when test data degradation matches the training ones, but they face challenges in generalizing to unknown degradations.
      To unleash the potential and generalization ability of SLMs, we propose a novel
      \textbf{self-supervised unknown-to-known degradation transformation framework (U2K)} for
      blind HSI fusion, which adaptively transforms unknown degradation into the same
      type of degradation as those handled by pre-trained SLMs. Specifically, the
      proposed U2K framework consists of: (1) spatial and spectral Degradation
      Wrapping (DW) modules that map HR-HSI to unknown degraded HR-MSI and LR-HSI,
      and (2) Degradation Transformation (DT) modules that convert these wrapped data
      into predefined degradation patterns. The transformed HR-MSI and LR-HSI pairs
      are then processed by a pre-trained network to reconstruct the target HR-HSI.
      We train the U2K framework in a self-supervised manner using consistency loss
      and greedy alternating optimization, significantly improving the flexibility of
      blind HSI fusion. 
      Extensive experiments confirm the effectiveness of our proposed U2K framework in \textbf{boosting the adaptability of five existing SLMs under various degradation settings} and surpassing state-of-the-art blind methods.
\end{abstract}

\section{Introduction}

\begin{figure}[t]
  \centering
  \includegraphics[height=185pt]{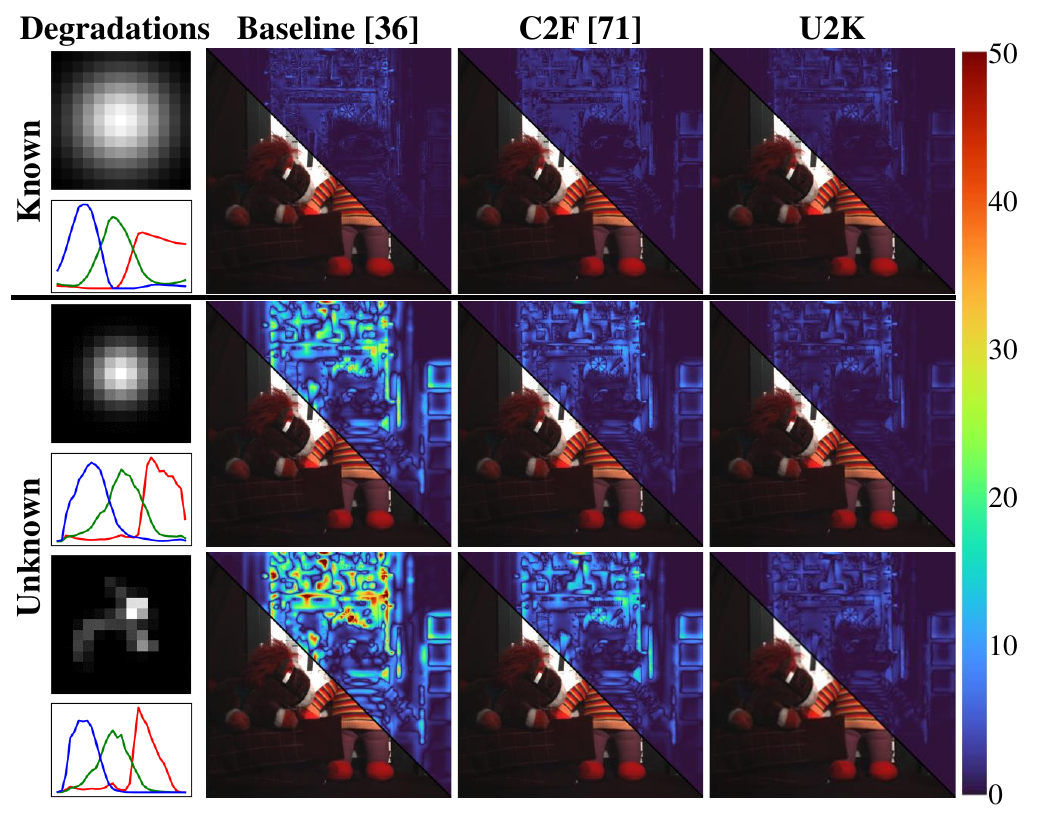}%
  \caption{
    Visual fusion results (bottom-left) and reconstruction error maps (top-right) under various degradation settings for the image \textit{chart\_and\_stuffed\_toy}. From left to right, the columns show the degradation operators and the results of the baseline \cite{ma2024reciprocal}, baseline with C2F \cite{zhang2024unsupervised}, and baseline with U2K, respectively. The baseline with U2K shows the best generalization capability. 
  }
  \label{fig:method_diff}
\end{figure}

Hyperspectral images (HSIs) capture both the spatial and spectral
characteristics of objects, enabling diverse applications in medical imaging
\cite{lu2014medical}, industrial inspection \cite{ma2019advanced}, and remote
sensing \cite{li2019deep}. However, inherent physical limitations impose a
trade-off between spatial and spectral resolution \cite{dian2021recent}. HSI
fusion has emerged as a promising solution to this challenge, aiming to
reconstruct the high-resolution HSI (HR-HSI)  by integrating a
low-resolution HSI (LR-HSI) with a high-resolution multispectral
image (HR-MSI).


The evolution of HSI fusion techniques can be divided into two main phases.
Early works primarily explore the intrinsic structure priors of HSIs, such as
sparsity \cite{akhtar2014sparse, xing2020joint}, low rank
\cite{yokoya2011coupled, zhang2016multispectral}, total variation
\cite{bungert2018blind, palsson2013new}, and non-local similarity
\cite{dian2019nonlocal, xu2019nonlocal}. Recently, supervised learning methods (SLMs) \cite{dong2021model, wang2021enhanced, dian2018deep, yang2018hyperspectral} have achieved unprecedented success by implicitly learning general HSI priors from large-scale datasets, significantly advancing the field.  However, as illustrated in \figref{fig:method_diff}, these methods
typically rely on predefined degradation models during training and demonstrate limited generalization to real-world scenarios with diverse and unknown degradations \cite{jung2020unsupervised, yang2023unsupervised}, greatly constraining their practical utility.

To this end, blind HSI fusion seeks to perform fusion without prior knowledge
of the degradations. Several methods \cite{yao2020cross, wang2020fusionnet,
      uezato2020guided, li2022deep} jointly estimate both the degradation and the
HR-HSI for each image pair 
in a self-supervised
manner. However, these approaches capture only partial structural priors of HSI
and are prone to local optima due to their image-dependent training. The
Coarse-to-Fine (C2F) strategy \cite{zhang2024unsupervised} refines the output of a pretrained fusion model at
inference time. As illustrated in \figref{fig:method_diff}, C2F effectively
corrects fusion results when unknown degradations exhibit minor deviations
from the known ones. However, \textbf{when degradations deviate significantly, C2F
fails to generalize, as error accumulation within the network further
exacerbates the deviation from the ground truth (GT)}.

To enhance the adaptability of existing networks while preserving supervised
image priors, we propose a novel Unknown-to-Known (U2K) framework, seamlessly
pluggable into any pretrained fusion network. 
Unlike C2F strategy, which refines fusion results during inference, U2K addresses deviations at the source, transforming unknown degradations into predefined patterns compatible with pretrained models and handling more extreme and inconsistent degradation scenarios
(\figref{fig:method_diff}). The U2K framework operates in a self-supervised
manner through two key components: (1) a Degradation Wrapping (DW) module,
which maps known HR-HSI to unknown degradation distributions, and (2) a
Degradation Transformation (DT) module, which converts wrapped data into
predefined degradations. The transformed data is processed by a pre-trained fusion network to reconstruct the original HR-HSI, accomplishing the self-supervised target. To prevent trivial solutions, we employ a consistency loss and optimize U2K using a Greedy Alternating Optimization (GAO) strategy.
In summary, our contributions are as follows:

\begin{itemize}
      \item We propose the first \textbf{U2K framework} for blind HSI fusion, which resolves unknown degradations at
            the source to align with SLMs, enhancing their adaptability to diverse
            degradations. 
      \item  We introduce two modules (\textbf{DW} and \textbf{DT}) and a
            \textbf{consistency loss}, optimized via \textbf{GAO} strategy, to
            enable effective self-supervised learning for U2K.
      \item  Extensive experiments demonstrate that U2K substantially improves the
            adaptability of existing SLMs and outperforms state-of-the-art (SOTA) blind fusion
            methods .
\end{itemize}

\section{Related Work}

\subsection{Model-Based HSI Fusion}
Model-based methods leverage physical imaging models to guide HSI fusion. For
instance, based on the spectral mixture model, LR-HSI and HR-MSI can be
decomposed into endmembers and abundances, allowing HR-HSI inference
\cite{akhtar2015bayesian}. Coupled matrix factorization
\cite{yokoya2011coupled} and nonnegative sparse representation
\cite{dong2016hyperspectral} further facilitate joint decomposition and
spectral dictionary learning in HSI fusion. Given the 3D nature of HSIs,
tensor-based methods, including coupled Tucker decomposition
\cite{li2018fusing}, tensor train \cite{dian2019learning}, canonical polyadic
\cite{xu2019nonlocal}, and tensor ring \cite{xu2020hyperspectral}, have been
explored. Bayes-based models incorporate degradation models and latent priors,
employing subspace sparsity \cite{wei2015hyperspectral}, group sparsity
\cite{han2018self}, low rank \cite{xue2021spatial}, manifold
\cite{zhang2018exploiting}, and total variation \cite{simoes2014convex}
regularizations. However, handcrafted priors often struggle to capture the
intricate spectral-spatial dependencies, ultimately constraining fusion
performance.
\subsection{Supervised HSI Fusion}
Supervised HSI fusion methods employ deep neural networks (DNNs) to learn mappings from observed data to HR-HSIs. Various architectures have been explored for this task,
including 2D convolutional neural networks (2D-CNNs) \cite{dian2018deep,
      yang2018hyperspectral, han2019multi, zhang2020ssr, xu2020ham,
      zhu2020hyperspectral, jiang2020learning, sun2021band, huJF2021hyperspectral,
      ran2023guidednet}, 3D-CNNs \cite{palsson2017multispectral, yang2020hybrid,
      li2021exploring, huJW2021hyperspectral}, and Transformers
\cite{hu2022fusformer, bandara2022hypertransformer, deng2023psrt,
      jia2023multiscale, wang2023mct, ma2024reciprocal, li2024casformer}. To enhance stability and interpretability, several approaches integrate degradation models and physical priors into deep learning frameworks by unfolding iterative optimization algorithms within network architectures \cite{shen2019spatial,
      xie2019multispectral, wei2020deep, dian2020regularizing, dong2021model,
      shen2021admm, huang2022deep}. Recently, diffusion models \cite{ho2020denoising,
      croitoru2023diffusion} have shown strong generative capabilities and achieved promising results in HSI fusion \cite{wu2023hsr, qu2024s2cyclediff,
      dong2024ispdiff}.
Although SLMs have significantly improved hyperspectral image (HSI) fusion, their effectiveness is limited by training data constraints, especially in real-world scenarios where unknown and highly variable degradations can severely degrade performance.
\subsection{Blind HSI Fusion}
Blind methods aim to fuse the LR-HSI and HR-MSI to obtain the target HR-HSI without prior knowledge of the degradations. The most straightforward
solution is to firstly estimate the spectral response function (SRF) and point spread function (PSF), then apply non-blind reconstruction methods \cite{wang2019deep}. However, this two-step strategy often fails to ensure optimal degradation parameters or reconstructed images. Recent methods \cite{yao2020cross, wang2020fusionnet, uezato2020guided, li2022deep} jointly estimate degradations and HR-HSI in a self-supervised manner for each image pair. Yet, per-pair training is computationally expensive and may lead to biased HSI structures due to limited data. To address this, Zhang et al. \cite{zhang2024unsupervised} proposed a C2F strategy integrated into SLMs, enabling adaptive blind fusion at inference time. However, C2F struggles with extreme degradation variations, limiting its practical applicability.

\section{The proposed U2K Framework}

\subsection{Motivation}

The relationship between the HR-HSI $\mathcal{Z}$ and its corresponding LR-HSI $\mathcal{X}$ and HR-MSI $\mathcal{Y}$ can be formally expressed through the following observation model:
\begin{align}
      \label{eq:observed}
      \begin{split}
            \mathcal{X} & = (\mathcal{Z} \ast k)_{\downarrow s} + \mathcal{N_X}, \\
            \mathcal{Y} & = \mathcal{Z} \times_3 R + \mathcal{N_Y},
      \end{split}
\end{align}
where $\mathcal{N_X}$ and $\mathcal{N_Y}$ represent additive noise terms, $k$ denotes the PSF and $R$ represents the SRF.

\begin{figure}[t]
    \centering 
        \includegraphics[height=150pt]{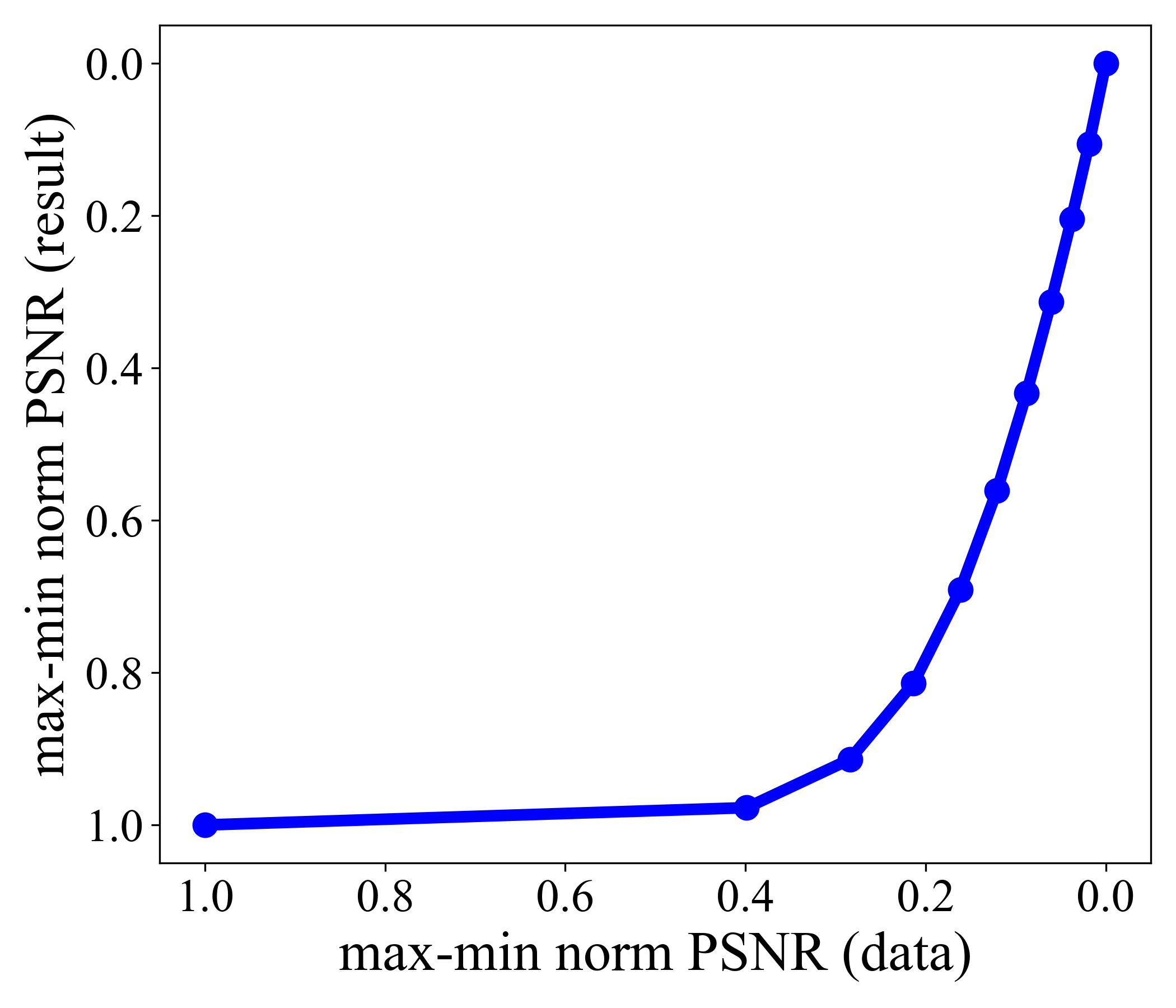}%
    \caption{
      The RSNR relationship 
      between the input data and the fusion results.
    }
    \label{fig:motivation}
\vspace{-10pt}
\end{figure}
\begin{figure*}[t]
  \centering
      \includegraphics[height=179pt]{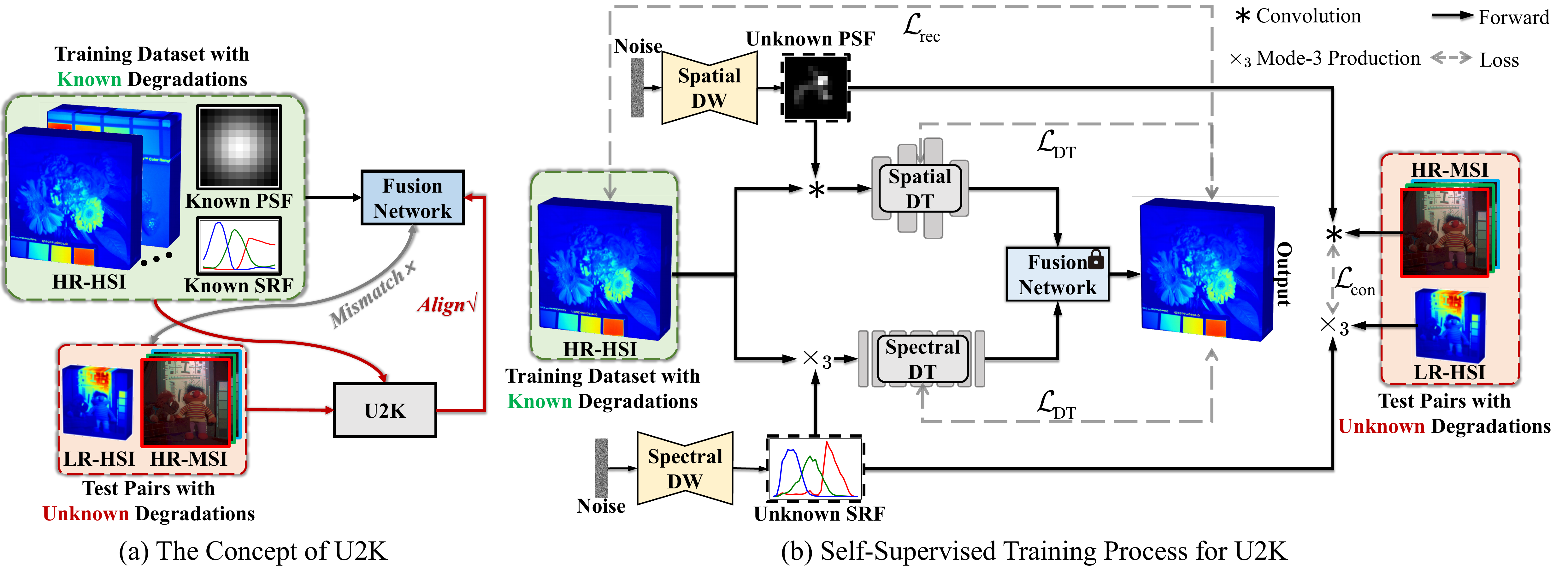}%
  \caption{
    The concept and training pipeline of our proposed U2K framework.
  }
  \label{fig:U2K}
\end{figure*}

A supervised fusion network, denoted as $\phi_F$, is typically trained on data generated by applying predefined degradations $(k, R)$ to HR-HSIs. While effective for known degradations, its generalization is inherently limited to real-world scenarios with unknown and various degradations. \textbf{A critical challenge lies in enabling $ \boldsymbol{\phi_F} $ to adapt to unknown degraded data without adjusting its parameters.} 
Before proceeding, we conduct a toy experiment on a pre-trained fusion network $\phi_F$ with different degradation settings. We plot the max-min normalized Peak Signal-to-Noise Ratio (PSNR)\footnote{Here, we just focus on the relative changes. The corresponding computational process is provided in \secref{sec:MMN_PSNR} of the Supplementary.} relationship between the tested data and the fusion results. As illustrated in \figref{fig:motivation}, $\phi_F$ exhibits limited adaptability to new degradations. Besides, the fusion quality deteriorates rapidly with increasing data distance due to the error accumulation within $\phi_F$. Considering this, post-processing corrections (like C2F) are challenging for large-deviated degradations. 
Also, \figref{fig:motivation} gives an inverse insight: minor adjustments of the input data can yield substantial performance improvements, providing a novel and critical perspective for effectively solving degradations with extreme deviations.
It motivates us to \textbf{address the problem at the source by establishing a mapping (\figref{fig:U2K}(a)) from unknown to known degradations,}
aligning the processed data within the capability of $\phi_F$.



Mathematically, given the test pair $(\mathcal{X}_{\text{test}}, \mathcal{Y}_{\text{test}})$ with unknown degradations $k_{\text{un}}$ and $R_{\text{un}}$,
\begin{align}
      \label{eq:new}
      \begin{split}
            \mathcal{X}_{\text{test}} = (\mathcal{Z}_{\text{test}} \ast k_{\text{un}})_{\downarrow s},   \mathcal{Y}_{\text{test}} = \mathcal{Z}_{\text{test}} \times_3 R_{\text{un}},
      \end{split}
\end{align}
we aim to learn a degradation transformation function $\mathcal{T}$ satisfying:
\begin{align}
      \label{eq:transformation}
      \begin{split}
            & \hat{\mathcal{X}}_{\text{test}}, \hat{\mathcal{Y}}_{\text{test}} = \mathcal{T}(\mathcal{X}_{\text{test}}, \mathcal{Y}_{\text{test}}), \\
            \text{subject to } & \hat{\mathcal{X}}_{\text{test}} = (\mathcal{Z}_{\text{test}} \ast k)_{\downarrow s}, \hat{\mathcal{Y}}_{\text{test}} = \mathcal{Z}_{\text{test}} \times_3 R.
      \end{split}
\end{align}
This transformation bridges the domain gap caused by unknown degradations, enabling $\phi_F$ to process new data effectively.

While supervised training of $\mathcal{T}$ would be straightforward given sufficient paired data $\{\mathcal{X}_{\text{test}}, \mathcal{Y}_{\text{test}}; \hat{\mathcal{X}}_{\text{test}}, \hat{\mathcal{Y}}_{\text{test}}\}$, such data is typically unavailable in practice. 
To overcome the limitation,  we propose the U2K framework, establishing a cyclic path between available HR-HSIs and unknown degradations, which learns $\mathcal{T}$  in a self-supervised manner. This approach enables $\phi_F$ to generate high-quality fusion results across diverse degradation types. The following sections detail the architecture and implementation of U2K.
 

\subsection{Overview}
\figref{fig:U2K} illustrates the conceptual framework and training pipeline of our proposed U2K. Specifically, as depicted in \figref{fig:U2K}(a), U2K is designed to mitigate the domain gap between the unknown degradation characteristics and the pretrained fusion network $\phi_F$, thereby addressing the inherent mismatch caused by unmodeled degradation patterns. In \figref{fig:U2K}(b), U2K implements a self-supervised cyclic learning process on available HR-HSIs through two core modules: 
the DW module and the DT module, each comprising specialized spatial and spectral components to handle respective degradations. 
The DW module distorts the HR-HSI $\mathcal{Z}$ to simulate unknown degradations, while the DT module maps the distorted observations back to the known degradation distribution. The outputs from the DT module are then fed into the pretrained fusion network $\phi_F$ to generate the final fused results. After optimization, the DT module and fusion path can be directly applied to new data with unknown degradations, generating high-quality fusion results\footnote{The inference process can be found in \secref{sec:inference} of the Supplementary.}.

\subsection{Degradation Wrapping}
The proposed DW module consists of two submodules: Spatial DW and Spectral DW. Spatial DW is responsible for spatially wrapping $\mathcal{Z}$ to coincide with the unknown degradation, and Spectral DW focuses on spectral adjustments.
\begin{align}
      \label{eq:degradation_wrapper}
      \begin{split}
            \tilde{\mathcal{X}} & = \text{SpectralDW}(\mathcal{Z}) \sim  \mathcal{X}_{\text{test}}, \\
            \tilde{\mathcal{Y}} & = \text{SpatialDW}(\mathcal{Z}) \sim  \mathcal{Y}_{\text{test}},
      \end{split}
\end{align}

To simplify the problem, we assume that the DW module itself is independent of the input image, meaning that the structure and parameters of DW remain consistent for any image. Replacing $\mathcal{Z}$ in
\eqref{eq:degradation_wrapper} with $\mathcal{Z}_{\text{test}}$ and considering
\eqref{eq:new} along with the designed purpose of DW, we have
$\text{SpectralDW}(\mathcal{Z}_{\text{test}}) = \mathcal{X}_{\text{test}}$ and
$\text{SpatialDW}(\mathcal{Z}_{\text{test}}) = \mathcal{Y}_{\text{test}}$. In
other words, with $\mathcal{Z}_{\text{test}}$ as input, DW essentially
identifies the unknown degradations. Based on the previous assumption, this is
also applicable to other $\mathcal{Z}$, allowing us to model the DW module in
the following form:
\begin{align}
      \label{eq:wrap}
      \begin{split}
            \text{SpectralDW}(\mathcal{Z}) & = \mathcal{Z} \times_3 \phi_R, \\
            \text{SpatialDW}(\mathcal{Z})  & =  (\mathcal{Z} \ast \phi_k)_{\downarrow s},
      \end{split}
\end{align}
where $\phi_R$ and $\phi_k$ are the parameterized degradations designed to approximate $k_{un}$ and $R_{un}$.
Unlike previous methods that employ a single convolution layer to represent the DW module \cite{zhang2020unsupervised}, our experiments reveal that a single convolution layer fails to optimize effectively for reasonable degradations. 
This is likely due to the high degree of freedom in the convolution kernel, making it prone to local minima. Furthermore, directly constraining the convolution kernel lacks flexibility. 
To address this, we adopt an MLP network to generate the degradations, with a Softmax operation to enforce non-negativity and a sum-to-one constraint.
Details on the optimization of the DW module are in \secref{section:loss_fun}.
\subsection{Degradation Transformation}
Similar to the DW module, the proposed DT module also includes two components:
spatial DT and spectral DT. However, the DT module has a more explicitly
defined output objective, which is to transform the distorted outputs
$\tilde{X}$ and $\tilde{Y}$ from the DW module into the known degraded forms
$\mathcal{X}$ and $\mathcal{Y}$.
\begin{align}
      \label{eq:DT}
      \begin{split}
            \hat{\mathcal{X}} & = \text{SpectralDT}(\tilde{X}) = \mathcal{X}, \\
            \hat{\mathcal{Y}} & = \text{SpatialDT}(\tilde{Y}) = \mathcal{Y}.
      \end{split}
\end{align}
$(\tilde{\mathcal{X}}, \tilde{\mathcal{Y}})$ and $(\mathcal{X}, \mathcal{Y})$ do not have a corresponding physical relationship and cannot be directly formulated into a degradation model.
Therefore, we employ a CNN with a residual structure to model the DT module, with appropriate adjustments for both Spatial and Spectral DT. 
The role of Spatial DT is to spatially aggregate pixels to adapt to $\mathcal{X}$; thus, we use a three-layer convolutional network with increasing receptive fields $(3,5,7)$ to approximate this process. 
On the other hand, spectral DT focuses on spectral reconstruction. Thus, we employ a six-layer $1 \times 1$ convolution with increasing channel dimensions for spectral aggregation.

\subsection{Loss Functions}
\label{section:loss_fun}

In this section, we will introduce the self-supervised loss used to train the
U2K framework and the consistency loss used to penalize the DW module.

\noindent\textbf{Self-supervised Loss}.
Our self-supervised loss consists of two parts. The first is the reconstruction loss of the U2K framework, where we use L1 loss to impose constraints between the input HR-HSI $\mathcal{Z}$ and the output of U2K $\hat{\mathcal{Z}}$:
\begin{align}
      \mathcal{L}_{rec} = || \mathcal{Z} - \hat{\mathcal{Z}}||_1.
\end{align}
Additionally, to optimize the DT module, we also use L1 loss to penalize the output of the DT module:
\begin{align}
      \mathcal{L}_{DT} = || \mathcal{X} - \hat{\mathcal{X}}||_1 + ||\mathcal{Y} - \hat{\mathcal{Y}}||_1.
\end{align}
We use a hyperparameter $\lambda$ to balance these two terms, so the self-supervised loss can be expressed as:
\begin{align}
      \mathcal{L}_{self} = \mathcal{L}_{rec} + \lambda\mathcal{L}_{DT}.
\end{align}
However, relying solely on the self-supervised loss is insufficient. When the DW module learns the
known degradations and the DT module collapses to an identity mapping, $\mathcal{L}_{self}$ may still approach 0.
To prevent this degenerate solution, we introduce an additional penalty on the DW module.


\noindent\textbf{Consistency Loss}.
As shown in \figref{fig:U2K}(b), feeding the spatially unknown degraded image
$\mathcal{X}_{\text{test}}$ into the Spectral DW module produces
$\mathcal{H}_{\mathcal{X}}$, which undergoes both unknown spatial and spectral
degradations. Similarly, we can obtain $\mathcal{H}_{\mathcal{Y}}$. Since
$\mathcal{H}_{\mathcal{X}}$ and $\mathcal{H}_{\mathcal{Y}}$ share a consistent
degradation pattern, we assume $\mathcal{H}_{\mathcal{X}} =
      \mathcal{H}_{\mathcal{Y}}$. Under this condition, we enforce an L1 loss
constraint to ensure consistency between Spectral DW and Spatial DW.
\begin{align}
      \label{eq:loss_con}
      \mathcal{L}_{con} = || \text{SpectralDT}(\mathcal{X}_{\text{test}}) - \text{SpatialDT}(\mathcal{Y}_{\text{test}})||_1.
\end{align}

Overall, the total loss function can be defined as:
\begin{align}
      \label{eq:loss}
      \mathcal{L} = \mathcal{L}_{\text{self}} + \mu\mathcal{L}_{\text{con}}.
\end{align}

\subsection{Greedy Alternation Optimization}

If gradient descent is applied directly to optimize all modules of the U2K framework jointly, the DW module may struggle to capture the intended degradations. We attribute this issue to misleading signals from the self-supervised loss \(\mathcal{L}_{self}\) during the early optimization stages.  
To address this challenge, we introduce a Greedy Alternating Optimization (GAO) strategy, where the DW and DT modules are updated alternately. 
Specifically, we perform multiple update steps on the DW module before each DT module update while reducing the learning rate of the DW module during DT optimization. 
This strategy mitigates the interference caused by initialization noise, which is common in naïve joint training. 
The optimization procedure of the U2K framework can be found in \secref{sec:GAO} of the Supplementary.

\section{Experiments}

\subsection{Experimental Settings}
\begin{figure}[t]
  \centering
      \includegraphics[height=139pt]{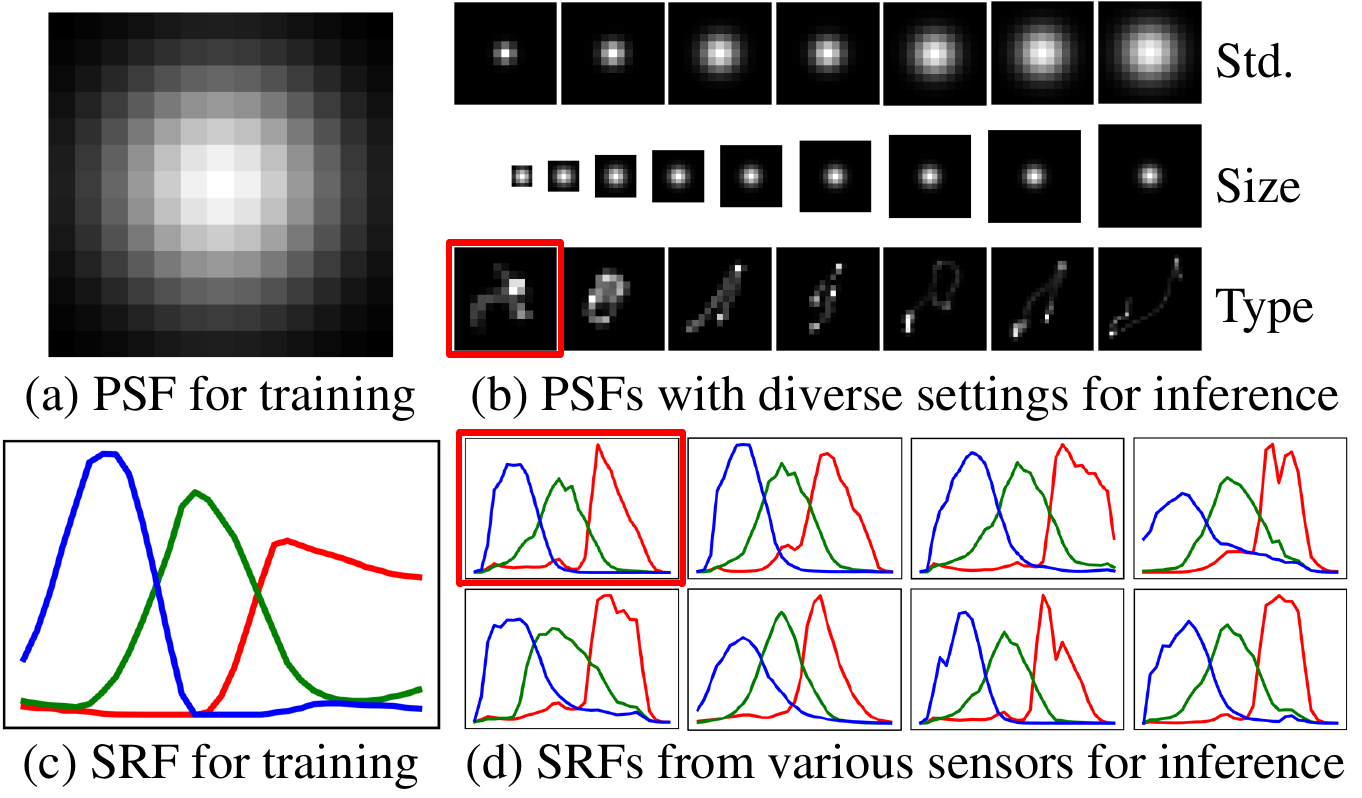}%
  \caption{
    Illustration of the degradations used in the experiments.
  }
  \label{fig:degradation}
    \vspace{-10pt}
\end{figure}

\textbf{Datasets.}
We evaluate our method on three synthetic HSI datasets (CAVE \cite{yasuma2010generalized}, Harvard \cite{chakrabarti2011statistics}, and Washington DC Mall (WDC) \cite{yokoya2017hyperspectral}) and three real-world datasets (two from GF1/5 and one from ZY1-02D).
Details are provided in \secref{sec:dataset} of the Supplementary.

\noindent\textbf{Dataset Division.}
For model training, we randomly select 16 and 25 HSIs from the CAVE and Harvard datasets, respectively. From each selected image, we extract overlapping patches of size $128 \times 128$ using a sliding window approach with a stride of 32. For the WDC dataset, we randomly partition the 16 cropped images into 8 for testing and the remaining 8 for training.

\noindent\textbf{Data Pair Generation.}
When generating HR-MSI and LR-HSI pairs, we apply distinct degradation processes to the training and testing datasets. 
For the training set, we simulate degradation using a $13\times13$ Gaussian kernel with a standard deviation ($\sigma$) of 3 and the SRF of a Nikon D700 camera\footnote{\url{https://maxmax.com/nikon_d700_study.htm}} . 
 
For the test set, we evaluate model robustness under more diverse conditions with (1) different Gaussian kernel sizes and $\sigma$ values, (2) different types of PSFs, and (3) SRFs from multiple sensors\footnote{\url{https://www.gujinwei.org/research/camspec/db.html}}.
The detailed degradation parameters for both datasets are visualized in \figref{fig:degradation}.

\noindent\textbf{Implementation Details.}
We implement our framework using PyTorch and conduct all experiments on an NVIDIA RTX 4090 GPU. The model is optimized using ADAM with $\beta_1 = 0.9$, $\beta_2 = 0.999$, and weight decay $\eta = 10^{-8}$. We set the learning rate to 0.001 for the spectral wrapper and 0.01 for other components. The training procedure consists of two phases: (1) 100 epochs for wrapper learning and (2) 10 epochs for adaptor learning, with a consistent batch size of 16 throughout the training process.

\subsection{Effectiveness of the U2K Framework}

\noindent\textbf{Adaptability to Unknown Degradations.}
When evaluating adaptability to unknown degradations, we first train DCTransformer \cite{ma2024reciprocal}, a SOTA supervised method, using the predefined degradation settings in \figref{fig:degradation}.
During testing, we assess the model performance under various degradation conditions. Using the unprocessed results as the baseline, we compare the performance of both C2F and U2K against this baseline, with quantitative results presented in \figref{fig:deg_setings}.

Our experiments reveal two key findings. 
Firstly, the baseline experiences a significant performance drop when the degradation settings in the test set deviate from those in the training set, with the severity of performance degradation proportional to the discrepancy.
This underscores the inherent limitation of supervised approaches in handling unseen degradation patterns. Secondly, while both C2F and U2K outperform the baseline across various degradation conditions, their robustness differs under large discrepancies, leading to notable performance divergence.
Specifically, C2F struggles to correct severely deviated fusion results, whereas U2K maintains remarkable stability. We attribute the superior robustness of U2K to its unique strategy of applying direct corrections at the source rather than the model result. This simplifies the correction process and enhances adaptability to diverse degradation scenarios.

\noindent\textbf{Adaptability to Fusion Networks.}
To comprehensively evaluate the versatility of the U2K framework, we integrate it with multiple SOTA fusion networks (denoted as method $A$). We establish three experimental configurations: (1) $A\text{-}S$: evaluation under degradations identical to the training set, (2) $A\text{-}D$: evaluation under degradations different from the training set, and (3) $A\text{-}D~\text{(U2K)}$: evaluation under different degradations with U2K framework integration. The comparison between $A\text{-}S$ and $A\text{-}D$ quantifies the transferability of method $A$ to unknown degradations, while the comparison between $A\text{-}D$ and $A\text{-}D~\text{(U2K)}$ validates the effectiveness of our framework.
We conduct extensive experiments with five representative supervised learning
models spanning two architectural paradigms: two CNN-based methods
(CNN \cite{dian2023zero} and SSR-Net \cite{zhang2020ssr}) and three
Transformer-based approaches (Fusformer \cite{hu2022fusformer},
SSTF-Unet \cite{liu2023sstf}, and MSST-Unet \cite{jia2023multiscale}).
\tabref{tab:e_U2K} presents the comprehensive quantitative results across
different downsampling scales. 
All methods exhibit substantial drops in performance when confronted with new degradations, with severity increasing with downsampling scale $s$ due to the heightened ill-posedness of \eqref{eq:new}. 
In contrast,
U2K consistently improves performance across all methods, demonstrating its robustness in improving network
generalization capabilities to handle unknown degradation scenarios. Besides,
while $A\text{-}D~\text{(U2K)}$  shows considerable gains, it still falls short of the ideal
$A\text{-}S$, suggesting promising directions for
further optimization within the U2K framework. 
Nevertheless, as shown in
the next section, the current U2K implementation achieves competitive
performance against SOTA blind methods,
underscoring its practical significance.

\subsection{Comparison with Blind Methods}
We select one model-based method CNMF 
\cite{yokoya2011coupled},
six deep learning approaches (including MHFNet \cite{xie2020mhf}, CUCaNet 
\cite{yao2020cross}, UAL  \cite{zhang2020unsupervised}, DEAM 
\cite{guo2023toward}, UTAL  \cite{zhang2024unsupervised}, UDTN \cite{wang2024unsupervised}) as comparisions.
All methods are trained on the dataset with predefined degradations and evaluated on the dataset with the degradations highlighted in red in \figref{fig:deg_setings}, using a downsampling factor of 32. In the following sections, unless stated otherwise, we adopt SSTF-Net as the fusion network for our U2K framework.

\tabref{tab:compare_BHSIF} aggregates the quantitative results on three simulated datasets, highlighting the superior performance of the proposed U2K. 
As illustrated in \figref{fig:results}, the proposed U2K method achieves the closest alignment with the GT and the minimum residual error across all datasets. It demonstrates that U2K consistently excels in capturing fine-grained details and preserving global contextual information, outperforming other methods in both qualitative and quantitative evaluations.

\subsection{Real-World Scene}
\begin{figure}[t]
    \centering
        \includegraphics[height=108pt]{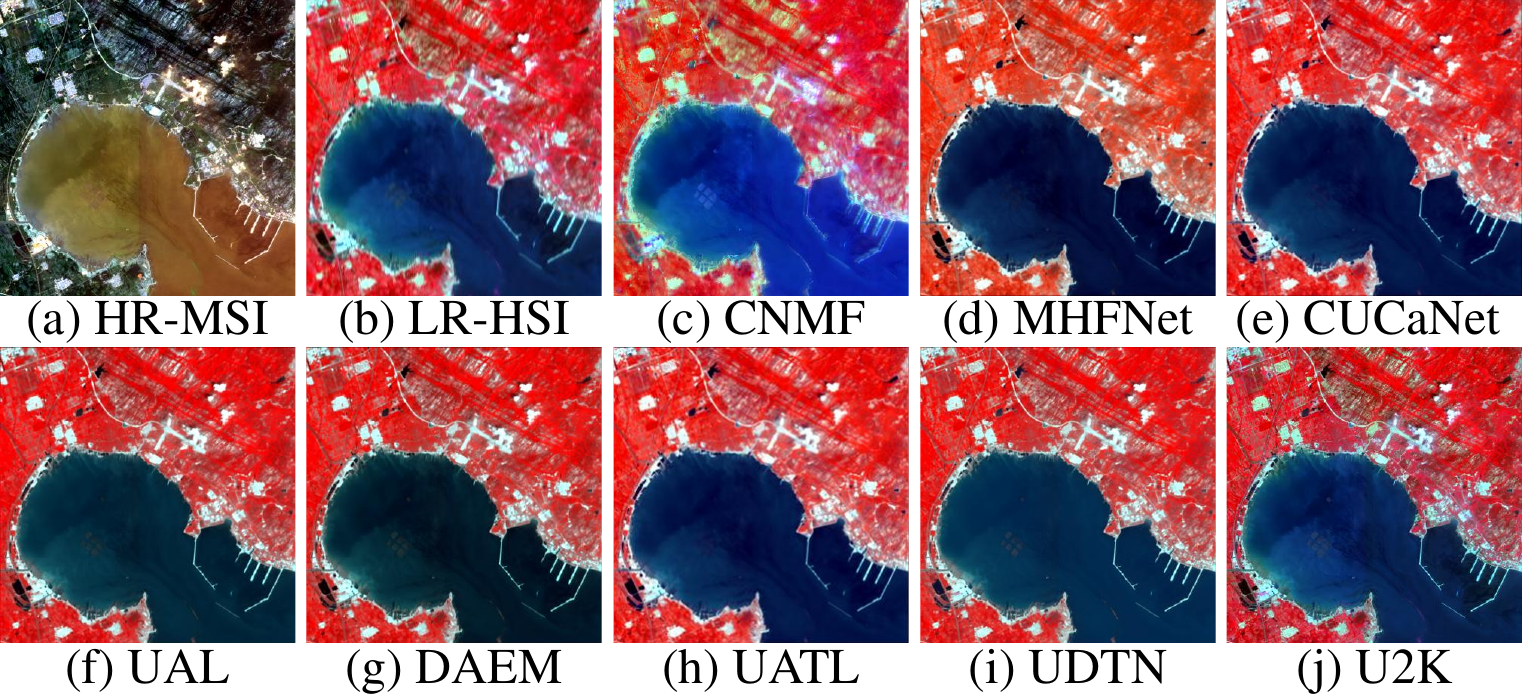}%
    \caption{
      Visual the HSI fusion result for the ZY1-02D satellite.
    }
    \label{fig:zy_small}
\end{figure}

We first present qualitative results on a real-world scene from the ZY-1D satellite in \figref{fig:zy_small}. Whereas existing blind methods often introduce spatial blurring or excessive spectral sharpening, U2K effectively preserves spatial details while maintaining high spectral fidelity, demonstrating its superior performance. The trade-off highlights the advantage of U2K in real-world applications. Due to space limitations, additional large-scale fusion results from the other two datasets are in \secref{sec:real-world} of the Supplementary.

\begin{figure}[t]
    \centering
        \includegraphics[height=80pt]{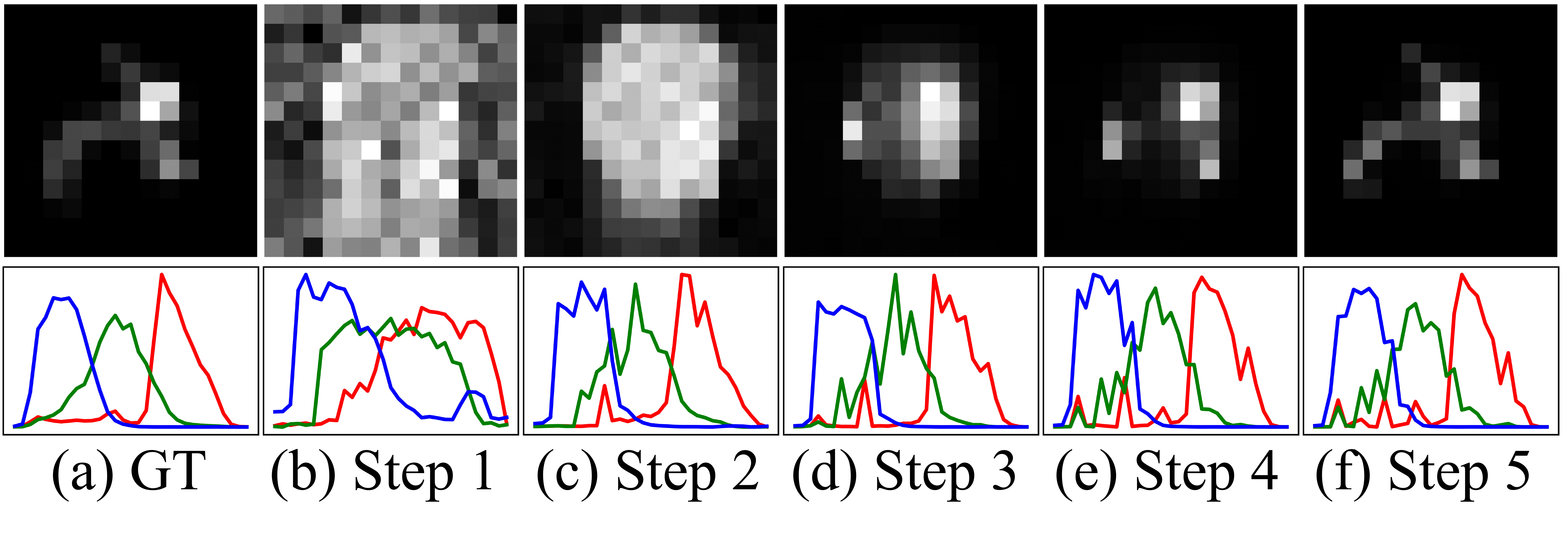}%
    \caption{
      The learned degradations from the DW module during the optimization process.
    }
    \label{fig:degradation_iter}
\end{figure}
\begin{figure}[t]
    \centering
        \includegraphics[height=132pt]{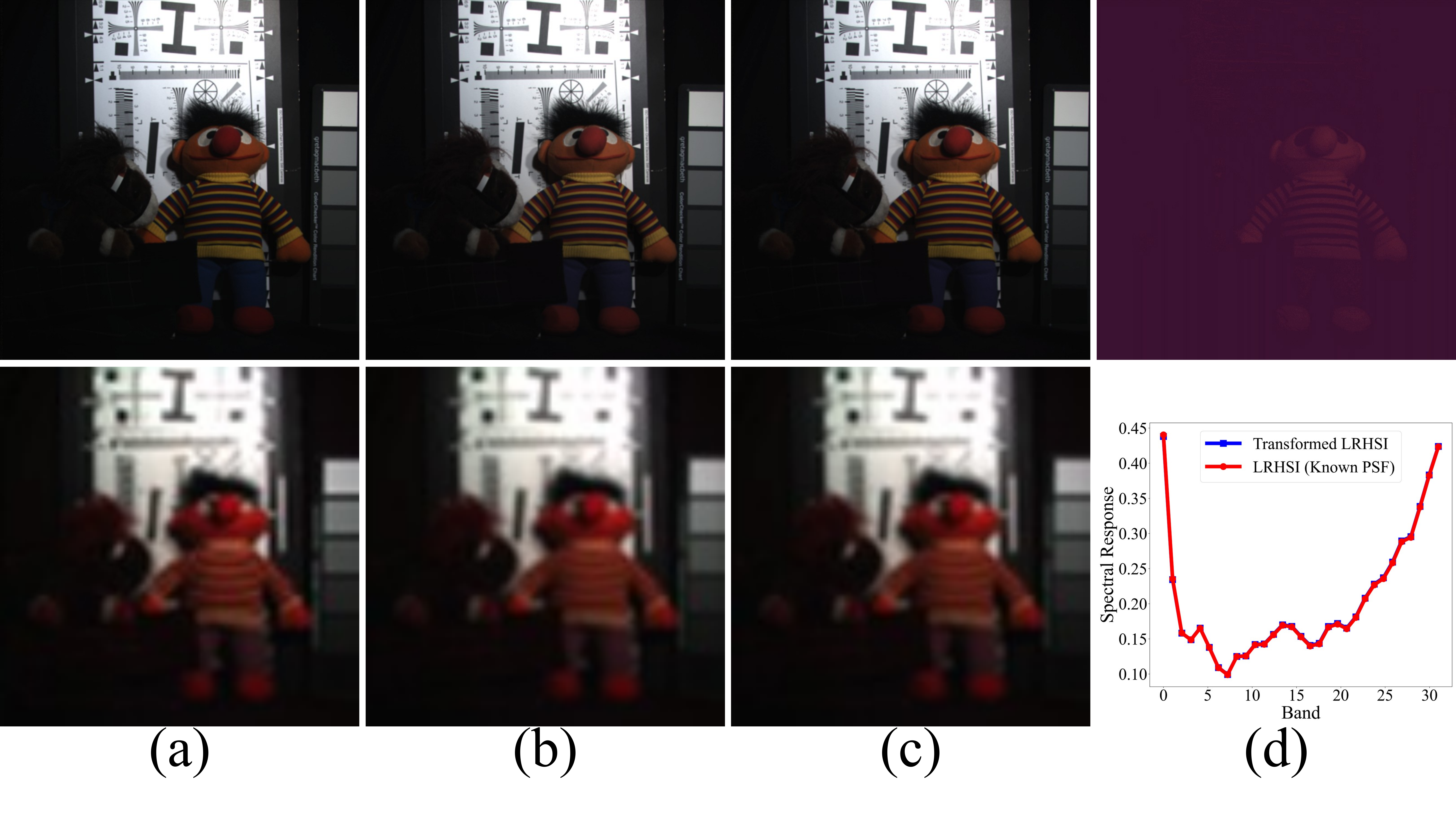}%
    \caption{
      The degraded (a) input, (b) output, (c) optimize-targeted HR-MSI (1st row) and LR-HSI (2nd row) pairs of the DT module, and (d) spatial and spectral differences between (b) and (c). 
    }
    \label{fig:inference}
    \vspace{-10pt}
\end{figure}

\subsection{Model Analysis}

\noindent\textbf{DW Module}. 
The proposed DW module distorts available HR-HSIs to fit the unknown degradation distribution. 
Under the assumption of image independence, it follows the form presented in \eqref{eq:wrap}, which estimates the unknown degradation. 
\figref{fig:degradation_iter} displays the PSF and SRF estimated by the DW during the optimization process.
Notably, the estimated PSF and SRF exhibit considerable noise in the early optimization stage, highlighting the necessity of adopting a greedy optimization approach for the DT module. 
 
As the number of iterations increases, the estimated results quickly converge toward the exact degradation, confirming the capability of the proposed DW module to learn the unknown degradation effectively.

\noindent\textbf{DT Module}. To illustrate the function of the DT module, we select a pair of $(\mathcal{X}_{\text{test}}, \mathcal{Y}_{\text{test}})$ from the test set and feed them into the DT module. Then, the results are compared with the corresponding known degraded image pairs, as shown in \figref{fig:inference}.
The transformation results of the DT module are very similar to the known degraded image pairs, indicating that DT can effectively convert unknown degraded images into a known distribution, making them suitable for use with existing fusion networks.

\subsection{Ablation Study}

\noindent\textbf{Loss Function Analysis.} As shown in \tabref{tab:loss},
the absence of the consistency loss $\mathcal{L}_{con}$ leads to model collapse into trivial solutions, causing severe performance degradation. The DT loss $\mathcal{L}_{DT}$ imposes a strict constraint on the DT module, requiring precise mapping from unknown degraded data to the known distribution. However, this strict constraint makes the model vulnerable to performance degradation from errors in individual modules. In contrast, the reconstruction loss $\mathcal{L}_{rec}$ enhances the adaptability of the U2K framework by relaxing the DT module requirements while considering the inherent adaptability of the fusion network.

\noindent\textbf{Optimization Strategy Comparison.}
We evaluate three optimization approaches: naive gradient descent with joint optimization, alternating optimization (AO), and our proposed GAO. As shown in \tabref{tab:GAO}, while AO improves model performance over joint optimization, it remains susceptible to noise during early optimization stages. Our GAO strategy addresses this limitation through a greedy approach, effectively mitigating noise impact and achieving superior performance.

\begin{table}
    \centering
    \small
    \setlength{\tabcolsep}{2pt}{
    \begin{tabular}{ccc|ccccc}
        \bottomrule[1.5pt]
        $\mathcal{L}_{rec}$ & $\mathcal{L}_{ET}$    & $\mathcal{L}_{con}$  & RMSE & PSNR  & SSIM  & SAM & ERGAS  \\
        \hline         
        \checkmark          & \checkmark            &                      & 0.1032	& 23.14	& 0.7826	& 29.32	& 18.7432  \\  
                            & \checkmark            & \checkmark           & 0.0187	& 38.06	& 0.9820	& 3.65	& 1.6832   \\ 
        \checkmark          &                       & \checkmark           & 0.0126	& 39.98	& 0.9852	& 2.92	& 1.5627   \\           
        \checkmark          & \checkmark            & \checkmark           & 0.0091	& 41.83	& 0.9889	& 2.15	& 1.3628   \\ 
        \toprule[1.5pt]
    \end{tabular}
    }
    \caption{
    Performance of  different losses.}
\label{tab:loss}
\end{table}

\begin{table}
    \centering
    \small
    \centering
    \setlength{\tabcolsep}{2pt}{
    \begin{tabular}{c|ccccc}
        \bottomrule[1.5pt]
        Optimization Method  & RMSE & PSNR  & SSIM  & SAM & ERGAS  \\
        \hline
        Naive                    & 0.0539	& 27.12	& 0.9246	& 9.53	& 4.7937 \\        
        AO                       & 0.0158	& 38.55	& 0.9831	& 3.29	& 1.6425   \\      
        GAO                      & 0.0091	& 41.83	& 0.9889	& 2.15	& 1.3628     \\    
        \toprule[1.5pt]
    \end{tabular}
    }
    \caption{
    Performance of different optimization strategies.}
\label{tab:GAO}
\vspace{-10pt}
\end{table}

\section{Conclusion}
 
To unleash the potential of existing supervised HSI fusion networks, we propose a novel U2K framework that transforms unknown degradations into known degradations while enabling seamless integration with current models.  
 
U2K incorporates the DW and DT modules for unknown-to-known degradation transformation learning. Furthermore, a consistency loss and GAO strategy are introduced and adopted for better optimization. Overall, U2K accomplishes self-supervised training. 
 
It excavates an efficient perspective for addressing unseen degradations and establishes a reliable path for transformation and fusion.
 
Experimental results show that U2K integrates well with fusion networks, effectively facilitating their adaptability and outperforming SOTA  blind methods, providing a powerful solution for HSI fusion.

      {
            \small
            \bibliographystyle{ieeenat_fullname}
            \bibliography{reference}

\begin{thebibliography}{73}
\providecommand{\natexlab}[1]{#1}
\providecommand{\url}[1]{\texttt{#1}}
\expandafter\ifx\csname urlstyle\endcsname\relax
  \providecommand{\doi}[1]{doi: #1}\else
  \providecommand{\doi}{doi: \begingroup \urlstyle{rm}\Url}\fi

\bibitem[Akhtar et~al.(2014)Akhtar, Shafait, and Mian]{akhtar2014sparse}
Naveed Akhtar, Faisal Shafait, and Ajmal Mian.
\newblock Sparse spatio-spectral representation for hyperspectral image
  super-resolution.
\newblock In \emph{Computer Vision--ECCV 2014: 13th European Conference,
  Zurich, Switzerland, September 6-12, 2014, Proceedings, Part VII 13}, pages
  63--78. Springer, 2014.

\bibitem[Akhtar et~al.(2015)Akhtar, Shafait, and Mian]{akhtar2015bayesian}
Naveed Akhtar, Faisal Shafait, and Ajmal Mian.
\newblock Bayesian sparse representation for hyperspectral image super
  resolution.
\newblock In \emph{Proceedings of the IEEE conference on computer vision and
  pattern recognition}, pages 3631--3640, 2015.

\bibitem[Bandara and Patel(2022)]{bandara2022hypertransformer}
Wele Gedara~Chaminda Bandara and Vishal~M Patel.
\newblock Hypertransformer: A textural and spectral feature fusion transformer
  for pansharpening.
\newblock In \emph{Proceedings of the IEEE/CVF conference on computer vision
  and pattern recognition}, pages 1767--1777, 2022.

\bibitem[Bungert et~al.(2018)Bungert, Coomes, Ehrhardt, Rasch, Reisenhofer, and
  Sch{\"o}nlieb]{bungert2018blind}
Leon Bungert, David~A Coomes, Matthias~J Ehrhardt, Jennifer Rasch, Rafael
  Reisenhofer, and Carola-Bibiane Sch{\"o}nlieb.
\newblock Blind image fusion for hyperspectral imaging with the directional
  total variation.
\newblock \emph{Inverse Problems}, 34\penalty0 (4):\penalty0 044003, 2018.

\bibitem[Chakrabarti and Zickler(2011)]{chakrabarti2011statistics}
Ayan Chakrabarti and Todd Zickler.
\newblock Statistics of real-world hyperspectral images.
\newblock In \emph{CVPR 2011}, pages 193--200. IEEE, 2011.

\bibitem[Croitoru et~al.(2023)Croitoru, Hondru, Ionescu, and
  Shah]{croitoru2023diffusion}
Florinel-Alin Croitoru, Vlad Hondru, Radu~Tudor Ionescu, and Mubarak Shah.
\newblock Diffusion models in vision: A survey.
\newblock \emph{IEEE Transactions on Pattern Analysis and Machine
  Intelligence}, 45\penalty0 (9):\penalty0 10850--10869, 2023.

\bibitem[Deng et~al.(2023)Deng, Deng, Wu, Ran, Hong, and Vivone]{deng2023psrt}
Shang-Qi Deng, Liang-Jian Deng, Xiao Wu, Ran Ran, Danfeng Hong, and Gemine
  Vivone.
\newblock Psrt: Pyramid shuffle-and-reshuffle transformer for multispectral and
  hyperspectral image fusion.
\newblock \emph{IEEE Transactions on Geoscience and Remote Sensing},
  61:\penalty0 1--15, 2023.

\bibitem[Dian et~al.(2018)Dian, Li, Guo, and Fang]{dian2018deep}
Renwei Dian, Shutao Li, Anjing Guo, and Leyuan Fang.
\newblock Deep hyperspectral image sharpening.
\newblock \emph{IEEE transactions on neural networks and learning systems},
  29\penalty0 (11):\penalty0 5345--5355, 2018.

\bibitem[Dian et~al.(2019{\natexlab{a}})Dian, Li, and Fang]{dian2019learning}
Renwei Dian, Shutao Li, and Leyuan Fang.
\newblock Learning a low tensor-train rank representation for hyperspectral
  image super-resolution.
\newblock \emph{IEEE transactions on neural networks and learning systems},
  30\penalty0 (9):\penalty0 2672--2683, 2019{\natexlab{a}}.

\bibitem[Dian et~al.(2019{\natexlab{b}})Dian, Li, Fang, Lu, and
  Bioucas-Dias]{dian2019nonlocal}
Renwei Dian, Shutao Li, Leyuan Fang, Ting Lu, and Jos{\'e}~M Bioucas-Dias.
\newblock Nonlocal sparse tensor factorization for semiblind hyperspectral and
  multispectral image fusion.
\newblock \emph{IEEE transactions on cybernetics}, 50\penalty0 (10):\penalty0
  4469--4480, 2019{\natexlab{b}}.

\bibitem[Dian et~al.(2020)Dian, Li, and Kang]{dian2020regularizing}
Renwei Dian, Shutao Li, and Xudong Kang.
\newblock Regularizing hyperspectral and multispectral image fusion by cnn
  denoiser.
\newblock \emph{IEEE transactions on neural networks and learning systems},
  32\penalty0 (3):\penalty0 1124--1135, 2020.

\bibitem[Dian et~al.(2021)Dian, Li, Sun, and Guo]{dian2021recent}
Renwei Dian, Shutao Li, Bin Sun, and Anjing Guo.
\newblock Recent advances and new guidelines on hyperspectral and multispectral
  image fusion.
\newblock \emph{Information Fusion}, 69:\penalty0 40--51, 2021.

\bibitem[Dian et~al.(2023)Dian, Guo, and Li]{dian2023zero}
Renwei Dian, Anjing Guo, and Shutao Li.
\newblock Zero-shot hyperspectral sharpening.
\newblock \emph{IEEE Transactions on Pattern Analysis and Machine
  Intelligence}, 45\penalty0 (10):\penalty0 12650--12666, 2023.

\bibitem[Dong et~al.(2016)Dong, Fu, Shi, Cao, Wu, Li, and
  Li]{dong2016hyperspectral}
Weisheng Dong, Fazuo Fu, Guangming Shi, Xun Cao, Jinjian Wu, Guangyu Li, and
  Xin Li.
\newblock Hyperspectral image super-resolution via non-negative structured
  sparse representation.
\newblock \emph{IEEE Transactions on Image Processing}, 25\penalty0
  (5):\penalty0 2337--2352, 2016.

\bibitem[Dong et~al.(2021)Dong, Zhou, Wu, Wu, Shi, and Li]{dong2021model}
Weisheng Dong, Chen Zhou, Fangfang Wu, Jinjian Wu, Guangming Shi, and Xin Li.
\newblock Model-guided deep hyperspectral image super-resolution.
\newblock \emph{IEEE Transactions on Image Processing}, 30:\penalty0
  5754--5768, 2021.

\bibitem[Dong et~al.(2024)Dong, Liu, Xiao, Qu, and Li]{dong2024ispdiff}
Wenqian Dong, Sen Liu, Song Xiao, Jiahui Qu, and Yunsong Li.
\newblock Ispdiff: Interpretable scale-propelled diffusion model for
  hyperspectral image super-resolution.
\newblock \emph{IEEE Transactions on Geoscience and Remote Sensing}, 2024.

\bibitem[Guo et~al.(2023)Guo, Xie, Jiang, Li, Lei, and Fang]{guo2023toward}
Wen-jin Guo, Weiying Xie, Kai Jiang, Yunsong Li, Jie Lei, and Leyuan Fang.
\newblock Toward stable, interpretable, and lightweight hyperspectral
  super-resolution.
\newblock In \emph{Proceedings of the IEEE/CVF Conference on Computer Vision
  and Pattern Recognition}, pages 22272--22281, 2023.

\bibitem[Han et~al.(2018)Han, Shi, and Zheng]{han2018self}
Xian-Hua Han, Boxin Shi, and Yinqiang Zheng.
\newblock Self-similarity constrained sparse representation for hyperspectral
  image super-resolution.
\newblock \emph{IEEE Transactions on Image Processing}, 27\penalty0
  (11):\penalty0 5625--5637, 2018.

\bibitem[Han et~al.(2019)Han, Zheng, and Chen]{han2019multi}
Xian-Hua Han, YinQiang Zheng, and Yen-Wei Chen.
\newblock Multi-level and multi-scale spatial and spectral fusion cnn for
  hyperspectral image super-resolution.
\newblock In \emph{Proceedings of the IEEE/CVF International Conference on
  Computer Vision Workshops}, pages 0--0, 2019.

\bibitem[Ho et~al.(2020)Ho, Jain, and Abbeel]{ho2020denoising}
Jonathan Ho, Ajay Jain, and Pieter Abbeel.
\newblock Denoising diffusion probabilistic models.
\newblock \emph{Advances in neural information processing systems},
  33:\penalty0 6840--6851, 2020.

\bibitem[Hu et~al.(2021{\natexlab{a}})Hu, Tang, Liu, and
  Fan]{huJW2021hyperspectral}
Jianwen Hu, Yuan Tang, Yaoting Liu, and Shaosheng Fan.
\newblock Hyperspectral image super-resolution based on multiscale mixed
  attention network fusion.
\newblock \emph{IEEE Geoscience and Remote Sensing Letters}, 19:\penalty0 1--5,
  2021{\natexlab{a}}.

\bibitem[Hu et~al.(2021{\natexlab{b}})Hu, Huang, Deng, Jiang, Vivone, and
  Chanussot]{huJF2021hyperspectral}
Jin-Fan Hu, Ting-Zhu Huang, Liang-Jian Deng, Tai-Xiang Jiang, Gemine Vivone,
  and Jocelyn Chanussot.
\newblock Hyperspectral image super-resolution via deep spatiospectral
  attention convolutional neural networks.
\newblock \emph{IEEE Transactions on Neural Networks and Learning Systems},
  33\penalty0 (12):\penalty0 7251--7265, 2021{\natexlab{b}}.

\bibitem[Hu et~al.(2022)Hu, Huang, Deng, Dou, Hong, and
  Vivone]{hu2022fusformer}
Jin-Fan Hu, Ting-Zhu Huang, Liang-Jian Deng, Hong-Xia Dou, Danfeng Hong, and
  Gemine Vivone.
\newblock Fusformer: A transformer-based fusion network for hyperspectral image
  super-resolution.
\newblock \emph{IEEE Geoscience and Remote Sensing Letters}, 19:\penalty0 1--5,
  2022.

\bibitem[Huang et~al.(2022)Huang, Dong, Wu, Li, Li, and Shi]{huang2022deep}
Tao Huang, Weisheng Dong, Jinjian Wu, Leida Li, Xin Li, and Guangming Shi.
\newblock Deep hyperspectral image fusion network with iterative
  spatio-spectral regularization.
\newblock \emph{IEEE Transactions on Computational Imaging}, 8:\penalty0
  201--214, 2022.

\bibitem[Jia et~al.(2023)Jia, Min, and Fu]{jia2023multiscale}
Sen Jia, Zhichao Min, and Xiyou Fu.
\newblock Multiscale spatial--spectral transformer network for hyperspectral
  and multispectral image fusion.
\newblock \emph{Information Fusion}, 96:\penalty0 117--129, 2023.

\bibitem[Jiang et~al.(2020)Jiang, Sun, Liu, and Ma]{jiang2020learning}
Junjun Jiang, He Sun, Xianming Liu, and Jiayi Ma.
\newblock Learning spatial-spectral prior for super-resolution of hyperspectral
  imagery.
\newblock \emph{IEEE Transactions on Computational Imaging}, 6:\penalty0
  1082--1096, 2020.

\bibitem[Jung et~al.(2020)Jung, Kim, Jang, Ha, and Sohn]{jung2020unsupervised}
Hyungjoo Jung, Youngjung Kim, Hyunsung Jang, Namkoo Ha, and Kwanghoon Sohn.
\newblock Unsupervised deep image fusion with structure tensor representations.
\newblock \emph{IEEE Transactions on Image Processing}, 29:\penalty0
  3845--3858, 2020.

\bibitem[Li et~al.(2024)Li, Zhang, Hong, Zhou, Vivone, Li, and
  Chanussot]{li2024casformer}
Chenyu Li, Bing Zhang, Danfeng Hong, Jun Zhou, Gemine Vivone, Shutao Li, and
  Jocelyn Chanussot.
\newblock Casformer: Cascaded transformers for fusion-aware computational
  hyperspectral imaging.
\newblock \emph{Information Fusion}, 108:\penalty0 102408, 2024.

\bibitem[Li et~al.(2022)Li, Zheng, Yao, Gao, and Hong]{li2022deep}
Jiaxin Li, Ke Zheng, Jing Yao, Lianru Gao, and Danfeng Hong.
\newblock Deep unsupervised blind hyperspectral and multispectral data fusion.
\newblock \emph{IEEE Geoscience and Remote Sensing Letters}, 19:\penalty0 1--5,
  2022.

\bibitem[Li et~al.(2021)Li, Wang, and Li]{li2021exploring}
Qiang Li, Qi Wang, and Xuelong Li.
\newblock Exploring the relationship between 2d/3d convolution for
  hyperspectral image super-resolution.
\newblock \emph{IEEE Transactions on Geoscience and Remote Sensing},
  59\penalty0 (10):\penalty0 8693--8703, 2021.

\bibitem[Li et~al.(2018)Li, Dian, Fang, and Bioucas-Dias]{li2018fusing}
Shutao Li, Renwei Dian, Leyuan Fang, and Jos{\'e}~M Bioucas-Dias.
\newblock Fusing hyperspectral and multispectral images via coupled sparse
  tensor factorization.
\newblock \emph{IEEE Transactions on Image Processing}, 27\penalty0
  (8):\penalty0 4118--4130, 2018.

\bibitem[Li et~al.(2019)Li, Song, Fang, Chen, Ghamisi, and
  Benediktsson]{li2019deep}
Shutao Li, Weiwei Song, Leyuan Fang, Yushi Chen, Pedram Ghamisi, and Jon~Atli
  Benediktsson.
\newblock Deep learning for hyperspectral image classification: An overview.
\newblock \emph{IEEE Transactions on Geoscience and Remote Sensing},
  57\penalty0 (9):\penalty0 6690--6709, 2019.

\bibitem[Liu et~al.(2023)Liu, Feng, Dian, and Li]{liu2023sstf}
Haibo Liu, Chenguo Feng, Renwei Dian, and Shutao Li.
\newblock Sstf-unet: Spatial--spectral transformer-based u-net for
  high-resolution hyperspectral image acquisition.
\newblock \emph{IEEE Transactions on Neural Networks and Learning Systems},
  2023.

\bibitem[Lu and Fei(2014)]{lu2014medical}
Guolan Lu and Baowei Fei.
\newblock Medical hyperspectral imaging: a review.
\newblock \emph{Journal of biomedical optics}, 19\penalty0 (1):\penalty0
  010901--010901, 2014.

\bibitem[Ma et~al.(2019)Ma, Sun, Pu, Cheng, and Wei]{ma2019advanced}
Ji Ma, Da-Wen Sun, Hongbin Pu, Jun-Hu Cheng, and Qingyi Wei.
\newblock Advanced techniques for hyperspectral imaging in the food industry:
  Principles and recent applications.
\newblock \emph{Annual review of food science and technology}, 10\penalty0
  (1):\penalty0 197--220, 2019.

\bibitem[Ma et~al.(2024)Ma, Jiang, Liu, and Ma]{ma2024reciprocal}
Qing Ma, Junjun Jiang, Xianming Liu, and Jiayi Ma.
\newblock Reciprocal transformer for hyperspectral and multispectral image
  fusion.
\newblock \emph{Information Fusion}, 104:\penalty0 102148, 2024.

\bibitem[Palsson et~al.(2013)Palsson, Sveinsson, and Ulfarsson]{palsson2013new}
Frosti Palsson, Johannes~R Sveinsson, and Magnus~O Ulfarsson.
\newblock A new pansharpening algorithm based on total variation.
\newblock \emph{IEEE Geoscience and Remote Sensing Letters}, 11\penalty0
  (1):\penalty0 318--322, 2013.

\bibitem[Palsson et~al.(2017)Palsson, Sveinsson, and
  Ulfarsson]{palsson2017multispectral}
Frosti Palsson, Johannes~R Sveinsson, and Magnus~O Ulfarsson.
\newblock Multispectral and hyperspectral image fusion using a
  3-d-convolutional neural network.
\newblock \emph{IEEE Geoscience and Remote Sensing Letters}, 14\penalty0
  (5):\penalty0 639--643, 2017.

\bibitem[Qu et~al.(2024)Qu, He, Dong, and Zhao]{qu2024s2cyclediff}
Jiahui Qu, Jie He, Wenqian Dong, and Jingyu Zhao.
\newblock S2cyclediff: Spatial-spectral-bilateral cycle-diffusion framework for
  hyperspectral image super-resolution.
\newblock In \emph{Proceedings of the AAAI Conference on Artificial
  Intelligence}, pages 4623--4631, 2024.

\bibitem[Ran et~al.(2023)Ran, Deng, Jiang, Hu, Chanussot, and
  Vivone]{ran2023guidednet}
Ran Ran, Liang-Jian Deng, Tai-Xiang Jiang, Jin-Fan Hu, Jocelyn Chanussot, and
  Gemine Vivone.
\newblock Guidednet: A general cnn fusion framework via high-resolution
  guidance for hyperspectral image super-resolution.
\newblock \emph{IEEE Transactions on Cybernetics}, 53\penalty0 (7):\penalty0
  4148--4161, 2023.

\bibitem[Shen et~al.(2021)Shen, Liu, Wu, Yang, and Xiao]{shen2021admm}
Dunbin Shen, Jianjun Liu, Zebin Wu, Jinlong Yang, and Liang Xiao.
\newblock Admm-hfnet: A matrix decomposition-based deep approach for
  hyperspectral image fusion.
\newblock \emph{IEEE Transactions on Geoscience and Remote Sensing},
  60:\penalty0 1--17, 2021.

\bibitem[Shen et~al.(2019)Shen, Jiang, Li, Yuan, Wei, and
  Zhang]{shen2019spatial}
Huanfeng Shen, Menghui Jiang, Jie Li, Qiangqiang Yuan, Yanchong Wei, and
  Liangpei Zhang.
\newblock Spatial--spectral fusion by combining deep learning and variational
  model.
\newblock \emph{IEEE Transactions on Geoscience and Remote Sensing},
  57\penalty0 (8):\penalty0 6169--6181, 2019.

\bibitem[Simoes et~al.(2014)Simoes, Bioucas-Dias, Almeida, and
  Chanussot]{simoes2014convex}
Miguel Simoes, Jos{\'e} Bioucas-Dias, Luis~B Almeida, and Jocelyn Chanussot.
\newblock A convex formulation for hyperspectral image superresolution via
  subspace-based regularization.
\newblock \emph{IEEE Transactions on Geoscience and Remote Sensing},
  53\penalty0 (6):\penalty0 3373--3388, 2014.

\bibitem[Sun et~al.(2021)Sun, Ren, Meng, Xiao, Yang, and Peng]{sun2021band}
Weiwei Sun, Kai Ren, Xiangchao Meng, Chenchao Xiao, Gang Yang, and Jiangtao
  Peng.
\newblock A band divide-and-conquer multispectral and hyperspectral image
  fusion method.
\newblock \emph{IEEE Transactions on Geoscience and Remote Sensing},
  60:\penalty0 1--13, 2021.

\bibitem[Uezato et~al.(2020)Uezato, Hong, Yokoya, and He]{uezato2020guided}
Tatsumi Uezato, Danfeng Hong, Naoto Yokoya, and Wei He.
\newblock Guided deep decoder: Unsupervised image pair fusion.
\newblock In \emph{European Conference on Computer Vision}, pages 87--102.
  Springer, 2020.

\bibitem[Wang et~al.(2024)Wang, Xu, Wu, and Wei]{wang2024unsupervised}
He Wang, Yang Xu, Zebin Wu, and Zhihui Wei.
\newblock Unsupervised hyperspectral and multispectral image blind fusion based
  on deep tucker decomposition network with spatial--spectral manifold
  learning.
\newblock \emph{IEEE Transactions on Neural Networks and Learning Systems},
  2024.

\bibitem[Wang et~al.(2019)Wang, Zeng, Huang, Ding, and Paisley]{wang2019deep}
Wu Wang, Weihong Zeng, Yue Huang, Xinghao Ding, and John Paisley.
\newblock Deep blind hyperspectral image fusion.
\newblock In \emph{Proceedings of the IEEE/CVF International Conference on
  Computer Vision}, pages 4150--4159, 2019.

\bibitem[Wang et~al.(2021)Wang, Fu, Zeng, Sun, Zhan, Huang, and
  Ding]{wang2021enhanced}
Wu Wang, Xueyang Fu, Weihong Zeng, Liyan Sun, Ronghui Zhan, Yue Huang, and
  Xinghao Ding.
\newblock Enhanced deep blind hyperspectral image fusion.
\newblock \emph{IEEE transactions on neural networks and learning systems},
  34\penalty0 (3):\penalty0 1513--1523, 2021.

\bibitem[Wang et~al.(2023)Wang, Wang, Song, Zhao, and Zhao]{wang2023mct}
Xianghai Wang, Xinying Wang, Ruoxi Song, Xiaoyang Zhao, and Keyun Zhao.
\newblock Mct-net: Multi-hierarchical cross transformer for hyperspectral and
  multispectral image fusion.
\newblock \emph{Knowledge-Based Systems}, 264:\penalty0 110362, 2023.

\bibitem[Wang et~al.(2020)Wang, Chen, Lu, Zhang, Liu, and
  Varshney]{wang2020fusionnet}
Zhengjue Wang, Bo Chen, Ruiying Lu, Hao Zhang, Hongwei Liu, and Pramod~K
  Varshney.
\newblock Fusionnet: An unsupervised convolutional variational network for
  hyperspectral and multispectral image fusion.
\newblock \emph{IEEE Transactions on Image Processing}, 29:\penalty0
  7565--7577, 2020.

\bibitem[Wei et~al.(2015)Wei, Bioucas-Dias, Dobigeon, and
  Tourneret]{wei2015hyperspectral}
Qi Wei, Jos{\'e} Bioucas-Dias, Nicolas Dobigeon, and Jean-Yves Tourneret.
\newblock Hyperspectral and multispectral image fusion based on a sparse
  representation.
\newblock \emph{IEEE Transactions on Geoscience and Remote Sensing},
  53\penalty0 (7):\penalty0 3658--3668, 2015.

\bibitem[Wei et~al.(2020)Wei, Nie, Li, Zhang, and Zhang]{wei2020deep}
Wei Wei, Jiangtao Nie, Yong Li, Lei Zhang, and Yanning Zhang.
\newblock Deep recursive network for hyperspectral image super-resolution.
\newblock \emph{IEEE Transactions on Computational Imaging}, 6:\penalty0
  1233--1244, 2020.

\bibitem[Wu et~al.(2023)Wu, Wang, Bai, Mao, Li, and Shen]{wu2023hsr}
Chanyue Wu, Dong Wang, Yunpeng Bai, Hanyu Mao, Ying Li, and Qiang Shen.
\newblock Hsr-diff: Hyperspectral image super-resolution via conditional
  diffusion models.
\newblock In \emph{Proceedings of the IEEE/CVF International Conference on
  Computer Vision}, pages 7083--7093, 2023.

\bibitem[Xie et~al.(2019)Xie, Zhou, Zhao, Meng, Zuo, and
  Xu]{xie2019multispectral}
Qi Xie, Minghao Zhou, Qian Zhao, Deyu Meng, Wangmeng Zuo, and Zongben Xu.
\newblock Multispectral and hyperspectral image fusion by ms/hs fusion net.
\newblock In \emph{Proceedings of the IEEE/CVF conference on computer vision
  and pattern recognition}, pages 1585--1594, 2019.

\bibitem[Xie et~al.(2020)Xie, Zhou, Zhao, Xu, and Meng]{xie2020mhf}
Qi Xie, Minghao Zhou, Qian Zhao, Zongben Xu, and Deyu Meng.
\newblock Mhf-net: An interpretable deep network for multispectral and
  hyperspectral image fusion.
\newblock \emph{IEEE Transactions on Pattern Analysis and Machine
  Intelligence}, 44\penalty0 (3):\penalty0 1457--1473, 2020.

\bibitem[Xing et~al.(2020)Xing, Wang, Dong, Duan, and Wang]{xing2020joint}
Changda Xing, Meiling Wang, Chong Dong, Chaowei Duan, and Zhisheng Wang.
\newblock Joint sparse-collaborative representation to fuse hyperspectral and
  multispectral images.
\newblock \emph{Signal Processing}, 173:\penalty0 107585, 2020.

\bibitem[Xu et~al.(2020{\natexlab{a}})Xu, Amira, Liu, Zhang, Zhang, and
  Li]{xu2020ham}
Shuang Xu, Ouafa Amira, Junmin Liu, Chun-Xia Zhang, Jiangshe Zhang, and
  Guanghai Li.
\newblock Ham-mfn: Hyperspectral and multispectral image multiscale fusion
  network with rap loss.
\newblock \emph{IEEE Transactions on Geoscience and Remote Sensing},
  58\penalty0 (7):\penalty0 4618--4628, 2020{\natexlab{a}}.

\bibitem[Xu et~al.(2019)Xu, Wu, Chanussot, Comon, and Wei]{xu2019nonlocal}
Yang Xu, Zebin Wu, Jocelyn Chanussot, Pierre Comon, and Zhihui Wei.
\newblock Nonlocal coupled tensor cp decomposition for hyperspectral and
  multispectral image fusion.
\newblock \emph{IEEE Transactions on Geoscience and Remote Sensing},
  58\penalty0 (1):\penalty0 348--362, 2019.

\bibitem[Xu et~al.(2020{\natexlab{b}})Xu, Wu, Chanussot, and
  Wei]{xu2020hyperspectral}
Yang Xu, Zebin Wu, Jocelyn Chanussot, and Zhihui Wei.
\newblock Hyperspectral images super-resolution via learning high-order coupled
  tensor ring representation.
\newblock \emph{IEEE transactions on neural networks and learning systems},
  31\penalty0 (11):\penalty0 4747--4760, 2020{\natexlab{b}}.

\bibitem[Xue et~al.(2021)Xue, Zhao, Bu, Liao, Chan, and
  Philips]{xue2021spatial}
Jize Xue, Yong-Qiang Zhao, Yuanyang Bu, Wenzhi Liao, Jonathan Cheung-Wai Chan,
  and Wilfried Philips.
\newblock Spatial-spectral structured sparse low-rank representation for
  hyperspectral image super-resolution.
\newblock \emph{IEEE Transactions on Image Processing}, 30:\penalty0
  3084--3097, 2021.

\bibitem[Yang et~al.(2018)Yang, Zhao, and Chan]{yang2018hyperspectral}
Jingxiang Yang, Yong-Qiang Zhao, and Jonathan Cheung-Wai Chan.
\newblock Hyperspectral and multispectral image fusion via deep two-branches
  convolutional neural network.
\newblock \emph{Remote Sensing}, 10\penalty0 (5):\penalty0 800, 2018.

\bibitem[Yang et~al.(2020)Yang, Xiao, Zhao, and Chan]{yang2020hybrid}
Jingxiang Yang, Liang Xiao, Yong-Qiang Zhao, and Jonathan Cheung-Wai Chan.
\newblock Hybrid local and nonlocal 3-d attentive cnn for hyperspectral image
  super-resolution.
\newblock \emph{IEEE Geoscience and Remote Sensing Letters}, 18\penalty0
  (7):\penalty0 1274--1278, 2020.

\bibitem[Yang et~al.(2023)Yang, Xiao, Zhao, and Chan]{yang2023unsupervised}
Jingxiang Yang, Liang Xiao, Yong-Qiang Zhao, and Jonathan Cheung-Wai Chan.
\newblock Unsupervised deep tensor network for hyperspectral--multispectral
  image fusion.
\newblock \emph{IEEE Transactions on Neural Networks and Learning Systems},
  2023.

\bibitem[Yao et~al.(2020)Yao, Hong, Chanussot, Meng, Zhu, and Xu]{yao2020cross}
Jing Yao, Danfeng Hong, Jocelyn Chanussot, Deyu Meng, Xiaoxiang Zhu, and
  Zongben Xu.
\newblock Cross-attention in coupled unmixing nets for unsupervised
  hyperspectral super-resolution.
\newblock In \emph{Computer Vision--ECCV 2020: 16th European Conference,
  Glasgow, UK, August 23--28, 2020, Proceedings, Part XXIX 16}, pages 208--224.
  Springer, 2020.

\bibitem[Yasuma et~al.(2010)Yasuma, Mitsunaga, Iso, and
  Nayar]{yasuma2010generalized}
Fumihito Yasuma, Tomoo Mitsunaga, Daisuke Iso, and Shree~K Nayar.
\newblock Generalized assorted pixel camera: postcapture control of resolution,
  dynamic range, and spectrum.
\newblock \emph{IEEE transactions on image processing}, 19\penalty0
  (9):\penalty0 2241--2253, 2010.

\bibitem[Yokoya et~al.(2011)Yokoya, Yairi, and Iwasaki]{yokoya2011coupled}
Naoto Yokoya, Takehisa Yairi, and Akira Iwasaki.
\newblock Coupled nonnegative matrix factorization unmixing for hyperspectral
  and multispectral data fusion.
\newblock \emph{IEEE Transactions on Geoscience and Remote Sensing},
  50\penalty0 (2):\penalty0 528--537, 2011.

\bibitem[Yokoya et~al.(2017)Yokoya, Grohnfeldt, and
  Chanussot]{yokoya2017hyperspectral}
Naoto Yokoya, Claas Grohnfeldt, and Jocelyn Chanussot.
\newblock Hyperspectral and multispectral data fusion: A comparative review of
  the recent literature.
\newblock \emph{IEEE Geoscience and Remote Sensing Magazine}, 5\penalty0
  (2):\penalty0 29--56, 2017.

\bibitem[Zhang et~al.(2016)Zhang, Wang, and Yang]{zhang2016multispectral}
Kai Zhang, Min Wang, and Shuyuan Yang.
\newblock Multispectral and hyperspectral image fusion based on group spectral
  embedding and low-rank factorization.
\newblock \emph{IEEE Transactions on Geoscience and Remote Sensing},
  55\penalty0 (3):\penalty0 1363--1371, 2016.

\bibitem[Zhang et~al.(2018)Zhang, Wei, Bai, Gao, and
  Zhang]{zhang2018exploiting}
Lei Zhang, Wei Wei, Chengcheng Bai, Yifan Gao, and Yanning Zhang.
\newblock Exploiting clustering manifold structure for hyperspectral imagery
  super-resolution.
\newblock \emph{IEEE Transactions on Image Processing}, 27\penalty0
  (12):\penalty0 5969--5982, 2018.

\bibitem[Zhang et~al.(2020{\natexlab{a}})Zhang, Nie, Wei, Zhang, Liao, and
  Shao]{zhang2020unsupervised}
Lei Zhang, Jiangtao Nie, Wei Wei, Yanning Zhang, Shengcai Liao, and Ling Shao.
\newblock Unsupervised adaptation learning for hyperspectral imagery
  super-resolution.
\newblock In \emph{Proceedings of the IEEE/CVF Conference on Computer Vision
  and Pattern Recognition}, pages 3073--3082, 2020{\natexlab{a}}.

\bibitem[Zhang et~al.(2024)Zhang, Nie, Wei, and Zhang]{zhang2024unsupervised}
Lei Zhang, Jiangtao Nie, Wei Wei, and Yanning Zhang.
\newblock Unsupervised test-time adaptation learning for effective
  hyperspectral image super-resolution with unknown degeneration.
\newblock \emph{IEEE Transactions on Pattern Analysis and Machine
  Intelligence}, 2024.

\bibitem[Zhang et~al.(2020{\natexlab{b}})Zhang, Huang, Wang, and
  Li]{zhang2020ssr}
Xueting Zhang, Wei Huang, Qi Wang, and Xuelong Li.
\newblock Ssr-net: Spatial--spectral reconstruction network for hyperspectral
  and multispectral image fusion.
\newblock \emph{IEEE Transactions on Geoscience and Remote Sensing},
  59\penalty0 (7):\penalty0 5953--5965, 2020{\natexlab{b}}.

\bibitem[Zhu et~al.(2020)Zhu, Hou, Chen, Zeng, and Zhou]{zhu2020hyperspectral}
Zhiyu Zhu, Junhui Hou, Jie Chen, Huanqiang Zeng, and Jiantao Zhou.
\newblock Hyperspectral image super-resolution via deep progressive
  zero-centric residual learning.
\newblock \emph{IEEE Transactions on Image Processing}, 30:\penalty0
  1423--1438, 2020.

\end{thebibliography}
      }

\end{document}